\documentclass[aps,prl,reprint,groupedaddress]{revtex4-1}

\bibstyle{apsrev4-1}

\usepackage{graphicx}
\usepackage{amssymb}
\usepackage{epstopdf}
\begin{document}

\preprint{}

\title{Statistical Physics of Self-Replication}


\author{Jeremy L. England}
\affiliation{Department of Physics, Massachusetts Institute of Technology, Building 6C, 77 Massachusetts Avenue, Cambridge, MA 02139}


\date{\today}

\begin{abstract}
Self-replication is a capacity common to every species of living thing, and simple physical intuition dictates that such a process must invariably be fueled by the production of entropy.  Here, we undertake to make this intuition rigorous and quantitative by deriving a lower bound for the amount of heat that is produced during a process of self-replication in a system coupled to a thermal bath.  We find that the minimum value for the physically allowed rate of heat production is determined by the growth rate, internal entropy, and durability of the replicator, and we discuss the implications of this finding for bacterial cell division, as well as for the pre-biotic emergence of self-replicating nucleic acids.
\end{abstract}

\pacs{}

\maketitle

\author{J. England}

\maketitle
Every living thing bears some resemblance to its ancestors; this is a basic premise of biology.  From the standpoint of physics, however, self-replication presents a challenge.  Reduced to its microscopic details, an organism cannot be distinguished from its environment: \emph{a priori}, a fluctuating cluster of atoms does not ``know" it is doing anything to affect the assembly of a similar-looking cluster.  As biologists, we watch a bacterial cell divide and say that it facilitates its own duplication, but as physicists, plotting the course from one microstate to the next, we cannot attribute any more agency to the atoms of the bacterium than we can to the atoms in the sugar it eats; all we see is interacting particles, exploring a series of arrangements permitted by conservation of momentum and energy.

     To resolve this difficulty, we must recognize that the ``self" in self-replication is not anywhere implicit in the atomistic physical description of the system.  Rather, it arises only once an observer carries out a further classification of microstates by providing a definition for some pattern of interest.  Such a coarse-graining of phase space should be familiar to any student of statistical mechanics, except that here, there need not be any way of physically summarizing (with, for example, an order parameter) the function of microscopic variables used to define the coarse-graining.

     To make things explicit, we might imagine a thought-experiment in which we showed every possible microstate for some system to a microbiologist who was asked to designate, in each case, how many live, healthy, wild-type \emph{E. coli} bacteria were present.  Though the microbiologist's assessment would be based partly or wholly on qualitative criteria, we would nevertheless come away from the procedure with a well-defined value for our cell count at each point in phase space.  In this way, holistic, biological judgments could be rendered into numbers that might then be incorporated into a quantitative model of the system's dynamics. 
    
         Here, we will use a coarse-graining like the one sketched above to investigate the statistical physics of self-replication.  Based on a general argument from microscopic reversibility, we will derive a lower bound on the heat output of a self-replicator in terms of its size, growth rate, entropy, and durability.  We will furthermore show, through analysis of empirical data, that this bound operates on a scale relevant to the functioning of real microorganisms and other self-replicators.
 
             We begin by considering the preparation of a large, finite system initially containing a single \emph{E. coli} bacterium, immersed in a sample of rich nutrient media in contact with a heat bath held at the bacterial cell's optimal growth temperature ($1/\beta \equiv T\sim 4.3\times 10^{-21}\textnormal{ joules}$) \cite{rothbaum,suckjoon}.  We can assume furthermore that the cell is in exponential growth phase at the beginning of its division cycle, and that, while the volume and mass of the entire system are held fixed, the composition and pressure of the nutrient media mimics that of a well-oxygenated sample open to the earth's atmosphere.  If we summarize the experimental conditions described above with the label $\mathbf{I}$, we can immediately say that there is some probability $p(i|\mathbf{I})$ that the system is found in some particular microstate $i$ given that it was prepared in the macroscopic condition $\mathbf{I}$ by some standard procedure.  Although this probability might well be impossible to derive \emph{ab initio}, in principle it could be measured through repeated consultations of a microbiologist as described above.
  
                Now suppose we consider what would happen in our system if we started off in some microstate $i$ and observed it again after a time interval of $\tau_{div}$, the typical duration of a single round of growth and cell division.  From the biological standpoint, the expected final state for the system is clear: two bacteria floating in the media instead of one, and various surrounding atoms rearranged into new molecular combinations (e.g. some oxygen converted into carbon dioxide).  While very likely, however, such an outcome is not certain, and in general we have to consider that each microstate $j$ will have some finite likelihood $p(j|\mathbf{II})$.  Since our system is coupled to a heat bath, it obeys stochastic dynamics described by the transition matrix $\pi(\rightarrow j |i)$, which is the conditional probability of ending up in microstate $j$ (with energy $E_{j}$) at time $t=\tau_{div}$ given that one started out in microstate $i$ (with energy $E_{i}$) at time $t=0$.  In these terms, our ensemble $\mathbf{II}$ of possible arrangements after time $\tau_{div}$ can be defined via
\begin{equation}	
p(j|\mathbf{II}) = \int_{\mathbf{I}} di~p(i|\mathbf{I})\pi(\rightarrow j|i),
\end{equation}
 which we take to have the normalization $\int dj~p(j|\mathbf{II})=1$.
 
By stipulation, there are no external driving forces acting on the system, and it can furthermore be assumed that the changes in our bacterial incubator over the course of time $\tau_{div}$ are dominated by diffusive motions that lack any sense of momentum \cite{crooks}.  Thus, the transition matrix $\pi(\rightarrow j |i)$ must obey a detailed balance relation
\begin{equation}\label{detbal}
\frac{\pi(\rightarrow i |j)}{\pi(\rightarrow j |i)} = \exp[-\beta \Delta Q_{i\rightarrow j}]
\end{equation}
where $\Delta Q_{i\rightarrow j}\equiv E_{i}-E_{j}$ is the heat released into the surrounding bath in the course of going from $i$ to $j$.  This follows from the underlying microscopic reversibility of particle dynamics in our system, along with the fact that the thermal equilibrium distribution $p(i)\propto \exp[-\beta E_{i}]$ must be stationary under $\pi(\rightarrow j|i)$ \cite{gardiner}.

To appreciate the thermodynamic consequences of the assumptions we have already made, it is necessary to consider the \emph{reverse} probability
\begin{equation}
\pi(\rightarrow \mathbf{I}|\mathbf{II}) = \int_{\mathbf{I}} di \int dj~p(j|\mathbf{II})\pi(\rightarrow i |j),
\end{equation}
that is, the minuscule probability that the system returns to one of the microstates satisfying the macroscopic condition $\mathbf{I}$ in time $\tau_{div}$ given that we start out with an initial distribution over microstates of $p(j|\mathbf{II})$.   Substituting from eq. (\ref{detbal}), we can rearrange this quantity to obtain
\begin{widetext}
\begin{equation}
\pi(\rightarrow \mathbf{I}|\mathbf{II}) =\int_{\mathbf{I}} di\int dj~\left(\frac{p(j|\mathbf{II})}{p(i|\mathbf{I})}\right) p(i|\mathbf{I})\pi(\rightarrow i |j)= \int_{\mathbf{I}} di\int dj~p(i|\mathbf{I})\pi(\rightarrow j |i) \left(\frac{e^{\ln \left[\frac{p(j|\mathbf{II})}{p(i|\mathbf{I})}\right]}}{e^{\beta \Delta Q_{i\rightarrow j}}}\right) =
\bigg\langle \frac{e^{\ln \left[\frac{p(j|\mathbf{II})}{p(i|\mathbf{I})}\right]}}{e^{\beta \Delta Q_{i\rightarrow j} }}\bigg\rangle_{\mathbf{I}\rightarrow \mathbf{II}}
\end{equation}
\end{widetext}
where $\langle\ldots\rangle_{\mathbf{I}\rightarrow \mathbf{II}}$ denotes an average over all paths from some $i$ in the initial ensemble $\mathbf{I}$ to some $j$ in the final ensemble $\mathbf{II}$, with each path weighted by its likelihood.

Defining the Shannon entropy $S$ for each ensemble in the usual manner ($S\equiv-\sum_{i}p_{i}\ln p_{i}$), we can construct $\Delta S_{int} \equiv S_{\mathbf{II}}-S_{\mathbf{I}}$, which measures the internal entropy change for the replication reaction.     Since it is generally the case that $ e^{x}\geq 1 +x$ for all $x$, we may rearrange (4) to write
\begin{equation}
\bigg\langle e^{-\beta \Delta Q_{i\rightarrow j}-\ln \pi(\rightarrow \mathbf{I}|\mathbf{II}) -\ln p(j|\mathbf{II})+\ln p(i|\mathbf{I}) }\bigg\rangle_{\mathbf{I}\rightarrow \mathbf{II}}=1
\end{equation}
and immediately arrive at
\begin{equation}\label{bound}
\beta\langle\Delta Q\rangle+\ln\left[\pi(\rightarrow\mathbf{I}| \mathbf{II})\right] +\Delta S_{int}\geq 0
\end{equation}
In one sense, this result simply says what we might have guessed: that the average total entropy production for the forward process $\langle\Delta S_{tot}\rangle\equiv\Delta S_{int} + \beta\langle\Delta Q\rangle$ sets a bound on how likely things would be to run in reverse: since the probability $\pi(\rightarrow\mathbf{I}|\mathbf{II}) \leq 1$, it follows that $\langle\Delta S_{tot}\rangle\geq 0$.  Put another way, what we have here is simply a precise statement of the Second Law of Thermodynamics.  It should therefore perhaps not be surprising to learn that the formula applies under conditions more general than those for which it was derived: in any diffusive, microscopically reversible system driven from equilibrium by a time-symmetric drive, it can be shown that $\pi(\rightarrow i|j)/\pi(\rightarrow j|i)=\langle \exp[-\beta \Delta Q_{i\rightarrow j}]\rangle_{i\rightarrow j}$ by using the irreversibility formula derived by Crooks \cite{crooks}.  Thus, although we have prefaced this investigation with a discussion of self-replication, our bound holds for a range of driven, nonequilibrium transitions between ensembles; in this light, we can see that the result is closely related to the well-known Landauer bound for the heat generated by the erasure of a bit of information \cite{landauer}.

Having established a general thermodynamic constraint on how self-replication can proceed, we now must consider whether or our finding is relevant in particular cases of interest. For the process of bacterial cell division introduced above, our ensemble $\mathbf{II}$ is a bath of nutrient-rich media containing two bacterial cells in exponential growth phase at the start of their division cycles.  In order to make use of the relation in (\ref{bound}), we need to estimate the likelihood that after time $\tau_{div}$, we will have ended up in an arrangement $\mathbf{I}$ where only one, newly formed bacterium is present in the system and another cell has somehow been converted back into the food from which it was built.  Our first hint that this likelihood must be quite small comes from our knowledge of the system's biology: if we start with two cells and wait a full division time, we will almost certainly end up with four! 
The challenge before us is therefore to quantify the extreme unlikelihood of the system doing something else. 

 The first piece is relatively easy to imagine: while we may not be able to compute the exact probability of a bacterium fluctuating to peptide-sized pieces and de-respirating a certain amount of carbon dioxide and water, we can be confident it is \emph{less} likely than all the peptide bonds in the bacterium spontaneously hydrolyzing.  Happily, this latter probability may be estimated in terms of the number of such bonds $n_{pep}$, the division time $\tau_{div}$, and the peptide bond half-life $\tau_{hyd}$. Assuming Poissonian statistics and a large value for $n_{pep}$, we have
$\ln p_{hyd}\simeq n_{pep}\ln[\tau_{div}/(n_{pep}\tau_{hyd})]$.

Handling the cell that stays alive is more challenging, as we have assumed this cell is growing processively, and we ought not make the mistake of thinking that such a reaction can be halted or paused (Fig. 1B) by a small perturbation.  The onset of exponential growth phase is preceded in \emph{E. coli} by a lag phase that can last several hours \cite{rothbaum}, during which gene expression is substantially altered so as to retool the cell for rapid division fueled by the available metabolic substrates \cite{lagexp}.  It is therefore appropriate to think of the cell in question as an optimized mixture of components primed to participate in irreversible, forward reactions like nutrient metabolism and protein synthesis.

 \begin{figure}
 \includegraphics[width=\columnwidth]{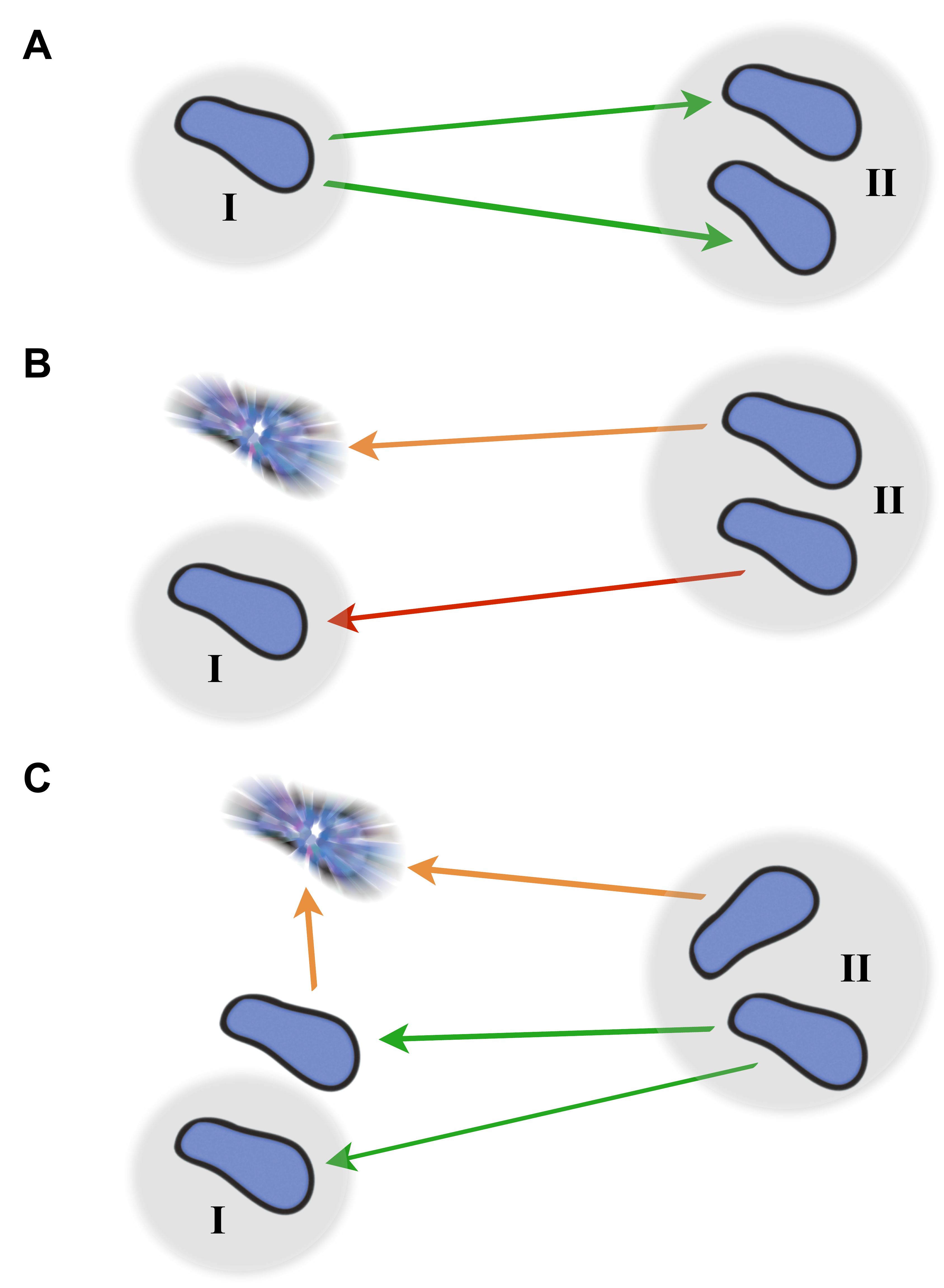}
 \caption{\label{fig}  A single bacterium in ensemble $\mathbf{I}$ is most likely to grow and divide (green arrows) as it progresses to ensemble $\mathbf{II}$ (panel A).  As panel B depicts, it is possible that a transition from $\mathbf{II}$ to $\mathbf{I}$ could be achieved by the spontaneous disintegration of one bacterial cell (orange arrows) accompanied by the spontaneous pausing of the growth of another one (red arrow).  However, it is arguable that the scenario displayed in panel C is more likely.  Here, one cell disintegrates, while another one divides into two cells, one of which then subsequently disintegrates. }
 \end{figure}

We can therefore argue that the likelihood of a spontaneous, sustained pause (of duration $\tau_{div}$) in the progression of these reactions is very small indeed: if each enzymatic protein component of the cell were to reject each attempt of a substrate to diffuse to its active site (assuming a diffusion time of small molecules between proteins of $\tau_{diff}\sim 10^{-8}\textnormal{ sec}$ \cite{protdiff,TIM}), we would expect  $|\ln p_{pause}|\propto |n_{pep}(\tau_{div}/\tau_{diff})|$ to exceed $|\ln p_{hyd}|$ by orders magnitude.  We must, however, consider an alternative mechanism for the most likely $\mathbf{II}\rightarrow \mathbf{I}$ transition (Fig. 1C): it is possible that a cell could grow and divide in an amount of time slightly less than $\tau_{div}$ \cite{suckjoon}.  If, subsequent to such an event, the daughter cell of the recent division were to spontaneously disintegrate back into its constituent nutrients (with log-probability at most on the order of $n_{pep}\ln[\tau_{div}/(n_{pep}\tau_{hyd})]$), we would complete the interval of $\tau_{div}$ with one, recently divided, processively growing bacterium in our system, that is, we would have returned to the $\mathbf{I}$ ensemble.  Thus, via a back-door into $\mathbf{I}$ provided to us by bacterial biology, we can claim that that 
\begin{equation}
\ln \pi(\rightarrow\mathbf{I} | \mathbf{II})\leq 2\ln p_{hyd} \simeq 2n_{pep}\ln[\tau_{div}/(n_{pep}\tau_{hyd})]
\end{equation}

Having obtained the above result, we can now refer back to the bound we set for the heat produced by this self-replication process and write 
\begin{equation}
\beta\langle Q\rangle \geq 2n_{pep}\ln[(n_{pep}\tau_{hyd})/\tau_{div}]-\Delta S_{int}
\end{equation}
This relation demonstrates that the heat evolved in the course of the cell making a copy of itself is set not only by the decrease in entropy required to arrange molecular components of the surrounding medium into a new organism, but also by how rapidly this takes place (through the division time $\tau_{div}$) and by how long we have to wait for the newly assembled structure to start falling apart (through $\tau_{hyd}$).  Moreover, we can now quantify the extent of each factor's contribution to the final outcome, in terms of $n_{pep}$, which we estimate to be $1.6\times 10^{9}$, assuming the dry mass of the bacterium is $0.3$ picograms \cite{biodata}. 

The total amount of heat produced in a single division cycle for an \emph{E. coli} bacterium growing at its maximum rate on lysogeny broth (a mixture of peptides and glucose) is $\beta\langle Q\rangle=220 n_{pep}$ \cite{rothbaum}.  We expect the largest contributions to the internal entropy change for cell division to come from the equimolar conversion of oxygen to carbon dioxide (since carbon dioxide has a significantly lower partial pressure in the atmosphere), and from the confinement of amino acids floating freely in the broth to specific locations inside bacterial proteins.  We can estimate the contribution of the first factor (which increases entropy) by noting that $\ln(\upsilon_{CO_{2}}/\upsilon_{O_{2}})\sim 6$.  The liberation of carbon from various metabolites also increases entropy by shuffling around vibrational and rotational degrees of freedom, but we only expect this to make some order unity modification to the entropy per carbon atom metabolized.  At the same time, peptide anabolism reduces entropy: by assuming that in 1\% tryptone broth, an amino acid starts with a volume to explore of $\upsilon_{i} = 100\textnormal{ nm}^{3}$ and ends up tightly folded up in some $\upsilon_{f} = 0.001 \textnormal{ nm}^{3}$ sub-volume of a protein, we obtain $\ln(\upsilon_{f}/\upsilon_{i}) \sim -12$.  In light of the fact that the bacterium consumes during division a number of oxygen molecules roughly equal to the number of amino acids in the new cell it creates \cite{rothbaum,cooney}, we can arbitrarily set a generous upper bound of $-\Delta S_{int}\leq 10 n_{pep}$.

In order to compare this contribution to that of the irreversibility term in (\ref{bound}), we assume a cell division time of $20$ minutes \cite{rothbaum,suckjoon}, and a spontaneous hydrolysis lifetime for peptide bonds of $600$ years at physiological pH \cite{peplife}, which yields $2n_{pep}\ln[\tau_{div}/(n_{pep}\tau_{hyd})] = 1.2 \times 10^{11}\simeq 75 n_{pep}$, a quantity at least several times larger than $\Delta S_{int}$.  Thus, as we may have expected based on previous evidence \cite{negentropy}, the entropic cost for aerobic bacterial respiration is relatively small, and is substantially outstripped by the sheer irreversibility of the self-replication reaction as it churns out copies that do not easily disintegrate into their constituent parts.

More significantly, these calculations also establish that the \emph{E. coli} bacterium produces an amount of heat less than three times as large as the absolute physical lower bound dictated by its growth rate, internal entropy production, and durability.  In light of the fact that the bacterium is a complex sensor of its environment that can very effectively adapt itself to growth in a broad range of different environments, we should not be surprised that it is not perfectly optimized for any given one of them.  Rather, it is remarkable that in a single environment, the organism can convert chemical energy into a new copy of itself so efficiently that if it were to produce even half as much heat it would be pushing the limits of what is thermodynamically possible!  This is especially the case since we deliberately underestimated the reverse reaction rate with our calculation of $p_{hyd}$, which does not account for the unlikelihood of spontaneously converting carbon dioxide back into oxygen.  Thus, a more accurate estimate of the lower bound on $\beta\langle Q\rangle$ in future may reveal \emph{E. coli} to be an even more exceptionally well-adapted self-replicator than it currently seems.

We have seen how this argument plays out for a bacterium, but the approach applies equally in a broad range of cases where there is some reliable stochastic model of a replicator's population dynamics \cite{gaspard}. For example, a recent study has used \emph{in vitro} evolution to optimize the growth rate of a self-replicating RNA molecule whose formation is accompanied by a single backbone ligation reaction and the leaving of a single pyrophosphate group \cite{RNArep}.  With a doubling time of $1$ hour, a half-life for RNA of $4$ years \cite{RNAstab}, and the reasonable assumption (in this case) that the change in translational entropy is negligible, we can estimate the heat bound as $\langle Q\rangle \geq RT\ln[(4 \textnormal{ years})/(1\textnormal{ hour})]= 7\textnormal{ kcal~mol}^{-1}$.  Since experimental data indicate an enthalpy for the reaction in the vicinity of $10\textnormal{ kcal~mol}^{-1}$ \cite{enthalpy,rnatherm}, it would seem this molecule operates quite near the limit of thermodynamic efficiency set by the way it is assembled.  

To underline this point, we may consider what the bound might be if this same reaction were somehow achieved using DNA, which is much more kinetically stable against hydrolysis than RNA \cite{DNAstab}.  In this case, we would have $\langle Q\rangle \geq RT\ln[(3\times10^{7} \textnormal{ years})/(1\textnormal{ hour})]= 16\textnormal{ kcal~mol}^{-1}$, which exceeds the estimated enthalpy for the ligation reaction and is therefore prohibited thermodynamically.   This calculation illustrates a significant difference between DNA and RNA, regarding each molecule's ability to participate in self-catalyzed replication reactions fueled by simple triphosphate building blocks: the far greater durability of DNA demands that a much higher per-base thermodynamic cost be paid in entropy production \cite{dzhang} in order for for the growth rate to match that of RNA in an all-things-equal comparison. Moreover, the heat bound difference between DNA and RNA should increase roughly linearly in $\ell$, the number of bases ligated during the reaction, which forces the maximum possible growth rate for a DNA replicator to shrink \emph{exponentially} with $\ell$ in comparison to that of its RNA equivalent.  This observation is certainly intriguing in light of past arguments made on other grounds that RNA, and not DNA, must have acted as the material for the pre-biotic emergence of self-replicating nucleic acids \cite{gilbert,rnatherm}.  

 The process of cellular division, even in a creature as ancient and streamlined as a bacterium, is so bewilderingly complex that it may come as some surprise that physics can make any binding pronouncements about how fast it all can happen.  The reason this becomes possible is that time-symmetrically driven, nonequilibrium processes in constant temperature baths obey general laws that relate forward and reverse transition probabilities to heat production \cite{crooks}.  Previously, such laws had been applied successfully in understanding thermodynamics of copying ``informational" molecules such as nucleic acids \cite{gaspard}.  In those cases, however, the information content of the system's molecular structure could more easily be taken for granted, in light of the clear role played by DNA in the production of RNA and protein.  What we have glimpsed here is that the underlying connection between entropy production and transition probability has a much more general applicability, so long as we recognize that ``self-replication" is something that happens relative to an observer: only once a classification scheme determines how many copies of some object are in the system for each microstate can we talk in probabilistic terms about the general tendency for that object to affect its own reproduction, and the same system's microstates can be classified using any number of different schemes.  We may hope that this insight spurs future work that will clarify the general physical constraints obeyed by natural selection in nonequilibrium systems. 

The author thanks C. Cooney,  J. Gore, A. Grosberg, D. Sivak, and A. Szabo for helpful comments.

\bibliography{mybib}

\end{document}